\documentclass{ws-ijmpa}

\begin{document}

\markboth{Mauricio Mondragon}
{COMMENT ON ``DIMENSION OF THE MODULI SPACE AND HAMILTONIAN ANALYSIS OF BF FIELD THEORIES''}

%
\catchline{}{}{}{}{}
%

\title{COMMENT ON ``DIMENSION OF THE MODULI SPACE AND HAMILTONIAN ANALYSIS OF BF FIELD THEORIES'' }

\author{MAURICIO MONDRAGON}

\address{D\'epartement de Langues \'etrang\`eres appliqu\'ees, Universit\'e de Provence, Centre Schuman,\\
29 avenue Robert Schuman, Aix-en-Provence 13621, France\\
momondragon@gmail.com}


\maketitle

\begin{history}
\received{Day Month Year}
\revised{Day Month Year}
\end{history}

\begin{abstract}
The purpose of this Comment is to point out that the results presented in the appendix of M.~Mondragon and M.~Montesinos, {\it J. Math. Phys.} {\bf 47}, 022301 (2006) provides a generic method so as to deal with cases as those of Section 6 of R. Cartas-Fuentevilla, A.~Escalante-Hern\'andez, and J.~Berra-Montiel, {\it Int. J. Mod. Phys. A} {\bf 26}, 3013 (2011). The results already reported are:  the canonical analysis, the transformations generated by the constraints, and the analysis of the reducibility of the
constraints for $SO(3,1)$ and $SO(4)$ four-dimensional BF theory coupled or not to a cosmological constant. But such
results are generic and hold actually for any Lie algebra having a non-degenerate inner product invariant under the action of the gauge group.

\end{abstract}

\ccode{PACS numbers: 04.20.Fy, 11.10.Ef, 03.50.-z}

The results reported in the appendix of Ref.~\refcite{jmp06} are generic enough so as to include findings like those presented by Cartas-Fuentevilla, Escalante-Hern\'andez,  and Berra-Montiel
in the Section 6 of Ref.~\refcite{otros}.

In fact, in Ref.~\refcite{jmp06} a detailed analysis of $SO(4)$ and $SO(3,1)$ BF theory with and without cosmological constant is carried out. Such an analysis is performed using the covariant canonical formalism as well as Dirac's canonical analysis. Furthermore, spacetime
diffeomorphisms and internal symmetries are analyzed too (on the issue of the symmetries, see also Ref.~ \refcite{cqg03}). One of the crucial results reported there is the analysis of the reducibility of some of the first-class constraints for $SO(3,1)$ and $SO(4)$ BF theory with and without cosmological constant.

The analysis reported in Ref.~\refcite{jmp06} is actually generic and holds for any $BF$ theory with and without cosmological constant. To appreciate this, consider the following action principle (see also Refs.~\refcite{cqg06} and \refcite{prd07})
\begin{eqnarray}\label{bfaction}
S[A,B] &=& \int_{M^4} \left [ B^i \wedge F_i -\frac12  \Lambda B^i \wedge B_i \right ],
\end{eqnarray}
where the $B$ field $B= B^i J_i$ is valued in the Lie algebra ${\mathfrak g}$ spanned by the generators $J^i$ and satisfy the commutation relations $[J_i, J_j]=C^k\,_{ij} J_k$ with $i,j,k,\dots = 1, \dots , dim({\mathfrak g})$. Also,  $F= F^i J_i$ is the curvature of the connection one-form $A=A^i J_i$ where $F^i=dA^i +
\frac12 C^i\,_{jk} A^j \wedge A^k$. The analysis is restricted to the case where the Lie algebra is endowed with a non-degenerate internal product invariant under the action of the internal group.

When the Dirac's canonical analysis of the action (\ref{bfaction}) is performed, one gets precisely the analogous results of those reported in the appendix of
Ref.~\refcite{jmp06}, but replacing in such formulae the pair of antisymmetric indices $IJ$ (associated with the Lie algebra of SO(4)) with the
index $i$ of this Comment (associated with the Lie algebra spanned by $J^i$). In particular, one gets (A7), (A8), (A9), and (A10), where now the constraints read
\begin{eqnarray}
{\tilde \Psi}^i = D_a \Pi^{ai} \approx 0, \quad {\tilde \Psi}^{ai} = \frac12 {\tilde \eta}^{abc} F^i_{bc}  - \Lambda \Pi^{ai} \approx 0
\end{eqnarray}
satisfying the following reducibility equations
\begin{eqnarray}\label{redu}
D_a {\tilde \Psi}^{ai} + \Lambda {\tilde \Psi}^i =0,
\end{eqnarray}
where $a,b,c =1,\dots 3$ are space indices. The counting of the degrees of freedom is as follows. There are $3 \times dim({\mathfrak g})$ configuration variables in $A^i_a$ and $3 \times dim({\mathfrak g})$ in their canonical momenta ${\tilde \Pi}^a_i$. There are
$dim({\mathfrak g})$ constraints in the Gauss law
${\tilde \Psi}^i$, $3 \times dim({\mathfrak g})$ first-class constraints in ${\tilde \Psi}^{ai}$, and there are $dim({\mathfrak g})$ reducibility
equations (\ref{redu}). Therefore, the counting gives $3 \times dim({\mathfrak g}) - (dim({\mathfrak g}) +
3 \times dim({\mathfrak g}) - dim({\mathfrak g})) =0$ local and physical degrees of freedom. The theory has therefore no physical excitations. To appreciate how gauge transformations and how diffeomorphisms arise see Ref.~\refcite{jmp06}.

Therefore, as the reader can see, the results reported in Ref.~\refcite{jmp06} are generic. There is nothing special in the specific internal Lie group or Lie algebra under consideration [$SO(3,1)$, $SO(4)$ or $SU(N)$ for example], what matters is that the Lie algebra is endowed with a non-degenerate inner product. Because of this, to make the analysis using one group or other is irrelevant. In particular, this shows that such analysis for the $SU(N)$ is potentially contained in Ref.~\refcite{jmp06} [confront the Eqs. (47), (50), (51), and (54) of Ref.~\refcite{otros} with the Eqs. (A7), (A8), (A9), and (A14) of Ref.~\refcite{jmp06}]. We wanted to rise this point for the sake of completeness.

\section*{Acknowledgments}
I am very greatful to Merced Montesinos for drawing my attention to this issue.

\end{document}